\documentclass[a4paper]{spie}  

\usepackage[]{graphicx}

\title{Deformations and Thermal Stability of Carbon Nanotube Ropes} 

\author
{Mar\'{\i}a J. L\'opez\supit{a}, Angel Rubio\supit{b}, and 
Julio A. Alonso\supit{c}
\skiplinehalf
\supit{a} Departamento de F\'{\i}sica Te\'orica, Universidad de Valladolid,
47011 Valladolid, Spain \\
\supit{b} Departamento de F\'{\i}sica de Materiales, Universidad del Pa\'{\i}s
Vasco and Donostia International Physics Center, DIPC,
20018 San Sebasti\'an/Donostia, Spain \\
\supit{c} Departamento F\'{\i}sica Te\'orica, Universidad de Valladolid,
47011 Valladolid, Spain and Donostia International Physics Center, DIPC,
20018 San Sebasti\'an/Donostia, Spain
}

\authorinfo{
Send correspondence to M.J.L.; E-mail: maria@lab2.fam.cie.uva.es
}
 
\begin{document} 
  \maketitle 

\begin{abstract}
Structural and thermal characteristics of crystalline ropes of 
single-wall carbon nanotubes (SWCNTs) are investigated. 
Novel crystalline ropes of polygonized SWCNTs produced by laser
irradiation exhibit rounded-hexagonal cross sections in contrast to 
earlier observations of circular tubes.
Extensive molecular dynamics (MD) simulations lead to several metastable 
structures of the lattice characterized by different tube cross sections, 
hexagonal, rounded-hexagonal and circular, and increasing cell volume. 
The competition between different tube shapes is analyzed and compared 
to experiments.
On the other hand, bundles of SWCNTs coalesce forming multiwall
carbon nanotubes (MWCNTs) under thermal treatment at high 
temperatures.
Extensive MD simulations confirm the SW-to-MW transformation
and suggest the physical patching--and--tearing mechanism
underlying the concerted coalescence of the tubes.
\end{abstract}

\keywords{carbon nanotubes, ropes, deformations, coalescence}

\section{INTRODUCTION}
\label{sect:intro}

As it is well known, carbon nanotubes (CNT) are seamless cylinders
of $sp^2$ carbon atoms arranged in a graphitic honeycomb structure.
Carbon nanotubes may be one fold (SWCNTs\cite{iij93}) or contain 
several cylinders nested one inside another (MWCNTs\cite{iij91}). 
SWCNTs, in many cases, self-organize into crystalline 
bundles\cite{the96,jou97} (set of a few to a few hundred aligned tubes 
arranged in a two-dimensional triangular lattice in the plane 
perpendicular to their common axes).
Although these metastable forms of carbon are subjected to intensive
investigations there are still important gaps in our understanding
and control of the properties of these materials.
A thorough understanding of carbon nanotubes includes the
detailed and comprehensive characterization of their possible
deformations due to interactions with a substrate or with other tubes,
and of their stability under thermal treatment or chemical agents.

It has been shown that CNTs may deform elastically away from their 
ideal circular cross sections when they interact either with a 
substrate \cite{her98} or with other tubes \cite{ruo93,ter94}.
Deformations of the tubes may have non trivial effects in their properties 
such as conductivity or phonon spectrum.
Tubes in a bundle interact one to another through attractive van der
Waals forces, similar to the ones acting between graphene layers
in graphite.
Therefore, the intertube interaction could prompt elastic structural 
changes in the tubes.
We have being able to synthesize novel crystalline bundles of 
``polygonized" (with hexagonal cross sections) SWCNT's \cite{lopez01}.
Our finding opens up the question of what is the equilibrium configuration
of a lattice of aligned tubes and the possibility of the existence of
several metastable structures which could be obtained depending on the
growth conditions.
To shed some light on this problem we have performed extensive molecular
dynamics simulations of lattices of monodisperse armchair and zigzag
SWCNTs as a function of tube diameter.
We find several metastable structures of the lattice characterized by
different tube cross sections, hexagonal, rounded-hexagonal and circular,
and increasing cell volume.
Our results presented in Section \ref{sect:polyg} help to interpret
both, the earlier (circular) and the presently (hexagonal) observed
tube shapes.

On the other hand, since CNTs are  metastable (the most stable form of
carbon is graphite), they may transform
into more stable structures under the appropriate annealing conditions.
We have found that bundles of SWCNTs coalesce forming MWCNTs, containing
from two to six nested tubes, under thermal treatment at high 
temperatures \cite{lamy98,metenier02,lopez02}.
This structural transformation is confirmed by extensive Molecular
Dynamics (MD) \cite{lopez02} simulations presented in 
Section \ref{sect:SWtoMW}.
The simulations combined with the experiments suggest
a ``patching--and--tearing" mechanism for the SW— to MWCNT's
transformation underlying the ``concerted" coalescence of the tubes
that begins with their polymerization.

\section{Theoretical Model}
\label{sect:thmod}

Due to the large number of atoms (up to 6000 nonequivalent atoms)
and the large time scales (of the order of 1000 ps)
involved in the simulation of the structural and thermal
properties of SWCNT bundles,
the use of accurate, ab initio quantum techniques in the 
description of these systems becomes impractical if not
completely unfeasible.
Therefore, we use a reliable and computational efficient, many-body 
interatomic potential to mimic the carbon-carbon interactions.
The potential consist in two terms:
i) a short-range part which is well suited for describing
the intratube covalent bonds \cite{ter88a,ter88b} and
ii) a long range term which describes the van de Waals interactions
between adjacent tubes \cite{nor96}.
This potential appropriately describes both, the covalent bond
in diamond and graphite and the van der Waals attraction between
graphene layers in graphite.
It has been successfully applied to the study of fullerenes and
tubes.

   \begin{figure}[b]
   \begin{center}
   \begin{tabular}{c}
   \includegraphics[height=8cm]{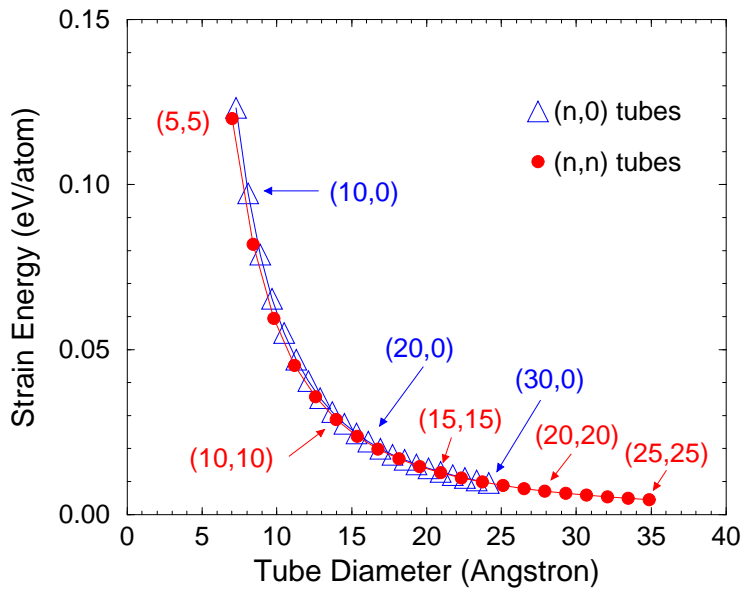}
   \end{tabular}
   \end{center}
   \caption
   { \label{figSW1} Strain energy of armchair (n,n) and zig-zag (n,0)
   SWCNTs as a function of tube diameter.}
   \end{figure}
The dynamical simulations are performed within the constant-energy
constant-volume ensemble.
The time evolution of the system is obtained by numerical integration,
using the velocity version of the Verlet algorithm,
of the classical Newtonian equations of motion.
To investigate possible metastable structures of the lattice of
SWCNTs a number of high energy configurations (for several lattice
parameters and various tube shapes) are generated and relaxed using
the thermal quenching procedure. 
This procedure consists in the step-wise removal of the kinetic 
energy of the system, along a dynamical trajectory, until the forces 
on all the atoms vanish and a local minimum on the potential energy 
surface has been found. 
The lowest energy minimum found will, likely, correspond to the
equilibrium configuration of the lattice and several metastable
structures will be also obtained.
We have also investigated the thermal behavior of bundles of
SWCNTs heat treated at high temperatures.
The simulation begins with cool tubes which are progressively heated 
up by scaling up the velocities of all the atoms.
The temperature of the bundles is raised up to a value 
(3000-3500$^{\rm o}$C) 
such that the rate of coalescence is high enough for being observed 
within the time scale ($t \leq 1000$ ps) of our simulations.

\section{Polygonized ropes}
\label{sect:polyg}

Novel crystalline ropes of polygonized SWCNTs have been produced by
CO$_2$ laser ablation\cite{lopez01}.
The ropes consist of monodisperse SWCNTs with  tube diameters $D$ of 
approximately 17 {\rm \AA} arranged in a triangular lattice.
HRTEM images of the lattice\cite{lopez01} clearly show the departure
of the tubes from the circular cross section and the appearance of facets,
parallel one to another, between adjacent tubes. 
The polygonized, hexagonal, cross section of the tubes is compatible
with the two-dimensional triangular symmetry of the lattice.
Our finding of ``hexagonal" tubes is in contrast to previous observations
of tubes with almost circular cross sections\cite{qin97}.
Then the question arises about the lowest energy structure of 
the lattice of aligned tubes and the possibility of having several 
metastable stuctures depending on the growth conditions.
To gain some insight into these problems we have performed an extensive 
search of possible metastable structures of the (triangular) lattice 
of aligned tubes.
Eventually, the equilibrium configuration of the lattice will be 
also found.

\vspace*{0.5 truecm}
   \begin{figure}[b]
   \begin{center}
   \begin{tabular}{c}
   \includegraphics[height=8cm]{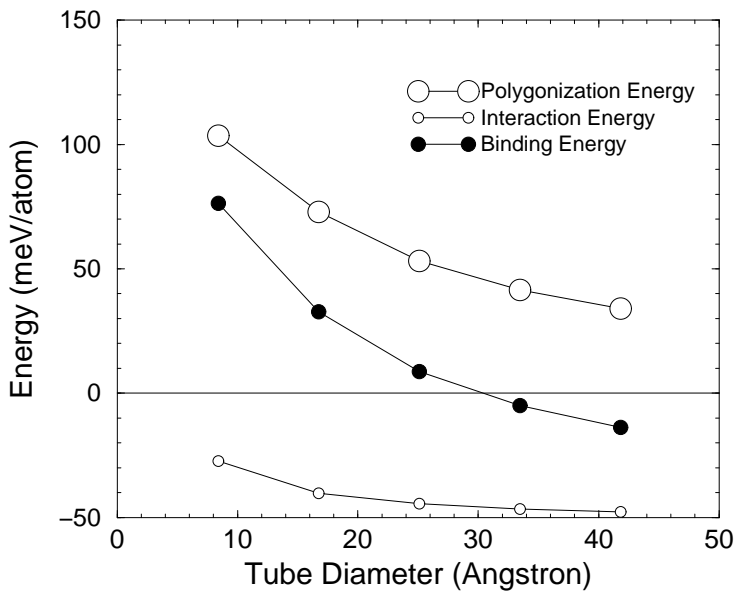}
   \end{tabular}
   \end{center}
   \caption
   {\label{figPOLYG1} Cohesive energy, with respect to the isolated 
   circular tubes, and intertube interaction energy of a lattice of 
   hexagonal (n,n) tubes as a function of tube diameter.
   Also shown the polygonization energy of the (n,n) hexagonal tubes.
   (See text for details).}
   \end{figure}
It is well established that free standing isolated tubes present circular
cross sections, since the circular shape minimizes the strain energy of 
the tubes. 
(In this context the strain energy is the elastic energy needed to roll 
up a graphene sheet into a cylinder to form a nanotube.)
Figure \ref{figSW1} shows the strain energy of armchair and zigzag SWCNTs 
as a function of tube diameter $D$.
This energy decreases as $1/R^2$ ($R$ is the tube radius) as expected
from the continium elastic model\cite{tibbetts84}.
The tubes, however, may deform elastically when they interact
either with a substrate or with other tubes.
Therefore, tubes in a bundle may depart, due to the intertube van der
Waals-type interaction, from the circular shape.
The deformation energy, i.e. energy cost of  changing the cross section
of the tubes from the ideal circular shape, should be, then, compensated
by the intertube interaction energy which acts as the driving force
for the elastic structural changes of the tubes.
Figure \ref{figPOLYG1} shows the deformation (polygonization) energy 
of armchair hexagonal tubes, the intertube interaction energy between
those tubes when forming a triangular lattice, and the cohesive 
energy of that  lattice of hexagonal tubes with respect to
the isolated circular tubes, all as functions of tube
diameter $D$.
The balance between the deformation energy, which opposes the 
polygonization of the tubes in the lattice, and the intertube interaction 
energy yields to stable lattices of hexagonal armchair tubes for
tube diameters larger than 30 {\rm \AA}.

   \begin{figure}[t]
   \begin{center}
   \begin{tabular}{c}
   \includegraphics[height=10cm]{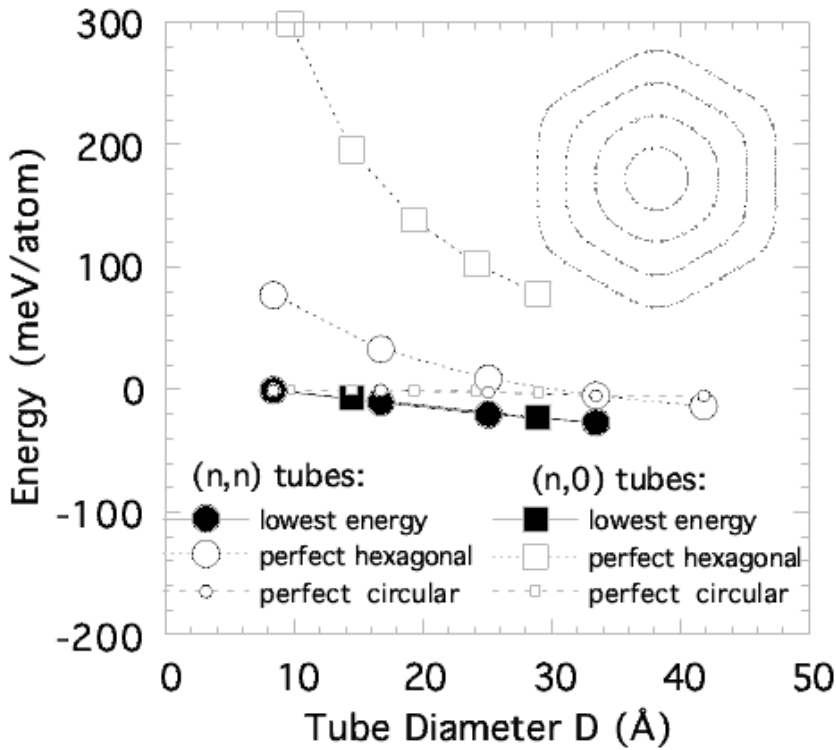}
   \end{tabular}
   \end{center}
   \caption
   {\label{figPOLYG2} Cohesive energy, with respect to the isolated 
   circular tubes, of the triangular lattice of 
   i) circular, ii) hexagonal, and iii) lowest energy tubes
   as a function of tube diameter.
   (n,n) armchair and (n,0) zigzag tubes are considered.
   The inset shows, superimposed, the structures of the lowest energy 
   (n,n) tubes.}
   \end{figure}
The lattice of circular tubes is always a bound system irrespective of
the diameter of the constituent tubes (see Fig. \ref{figPOLYG2}).
Its cohesive energy, however, is pretty small as compared
to the interaction energy between graphene layers in graphite.
The reason being in the lack of planar interacting faces
between adjacent tubes. 
Notice that the interaction energy between hexagonal tubes, which exhibit
planar facets lying parallel one to another between adjacent tubes,
is 1 order of magnitude larger than that of the lattice of circular tubes
(see Fig. \ref{figPOLYG1}).
However, for small diameter tubes, this energy does not compensate the
high deformation energy of hexagonal tubes and the lattice of circular 
tubes is more stable than the corresponding lattice of hexagonal tubes.
Eventually, since the deformation energy decreases with increasing 
tube diameter, the lattice of hexagonal tubes becomes more stable 
than the lattice of circular tubes (see Fig. \ref{figPOLYG2}).
The onset of polygonization, $D$=34 and 50 {\rm \AA} for armchair 
and zigzag tubes, respectively, is, however, too high as compared with
the experimental value, $D$=17 {\rm \AA}.

To obtain metastable, and eventually the lowest energy, structures of
the lattice of tubes we have applied the thermal quenching procedure.
First, a number of lattice configurations corresponding to several 
values of the lattice parameter and to tube shapes
ranging from circular to hexagonal are generated.
Then, all those high energy configurations of the lattice are cool down 
into rigid structures corresponding to local minima of the potential
energy surface.
The relaxations yield several metastable structures characterized by 
different tube cross sections and unit cell volume.
The shape of the tubes changes, with increasing cell volume, from
``nearly" hexagonal to ``almost" circular going through an
intermediate volume lattice formed by hexagonal tubes with rounded 
corners.
The lowest energy configuration of the lattice evolves from
{\em circular} to {\em hexagonal} tubes with increasing tube diameter.
For $D\leq 17$ {\rm \AA} nearly circular tubes are found.
This explains why all previous observations of bundles of tubes in 
this size range show tubes with circular cross sections.

   \begin{figure}[t]
   \begin{center}
   \begin{tabular}{c}
   \includegraphics[height=6cm]{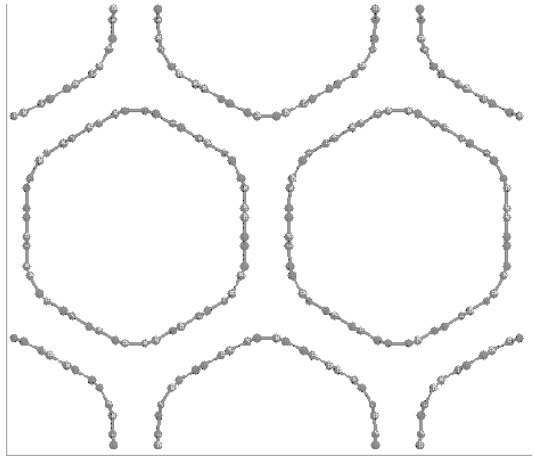}
   \end{tabular}
   \end{center}
   \caption
   {\label{figPOLYG3} Metastable structure of a lattice of (12,12) tubes
   with rounded hexagonal cross section. This structure supports
   the novel crystalline ropes of  polygonized tubes produced by CO$_2$
   laser ablation.}
   \end{figure}
However, the metastable structure, shown in Figure \ref{figPOLYG3},
formed by rounded hexagonal tubes
is only 0.6 meV/atom above the circular lowest energy structure of
(12,12) tubes.
Similarly the metastable structure of rounded hexagonal (18,0) tubes is
only 2.8 meV/atom above the minimum.
These novel metastable structures give support to the crystalline 
bundles of polygonized tubes found in the experiment.

\section{Thermal transformations of SWCNT ropes}
\label{sect:SWtoMW}

Coalescence of two SWCNTs into a larger diameter SWCNT has been observed 
upon the annealing of SWCNTs at high temperatures either in the presence
of H$_2$\cite{nik97} or under electron irradiation\cite{terr00}.
We have discovered\cite{lopez02}, as we were investigating the stability 
of SWCNT ropes under thermal treatment, a new structural transformation
of bundles of SWCNTs into MWCNTs.
Samples 
\footnote{
HRTEM characterization of the samples shows that they are made of bundles of
SWCNTs with bundle diameters ranging from 5 to 20~nm.
The individual tubes of about 1.4~nm in diameter are arranged in a
triangular lattice.
} 
of SWCNT ropes produced by the catalytic arc--discharge technique\cite{jou97}
were heated for about 15 min. at temperatures between 1600--2800$^{\rm o}$C 
under argon flow.
Above 2200$^{\rm o}$C the bundles disappear giving rise to MWCNTs 
consisting of two to six nested tubes.
All the characterization techniques of the samples, i.e., 
High Resolution Transmission Electron Microscopy (HRTEM), X-ray 
diffraction and Raman spectroscopy, confirm the structural reorganization 
of the bundles.
This SW--to--MW transformation is also confirmed by extensive Molecular 
Dynamics (MD) simulations (see below).
The simulations combined with the experimental results unveil the 
physical ``patching--and--tearing" mechanism of this showy transformation. 
This novel mechanism is of general applicability and describes both 
the coalescence of two SWCNTs into a larger diameter SWCNT and
the newest transformation of SWCNT ropes into MWCNTs.

\subsection{Coalescence of two tubes}
\label{subsect:coalescence}

First we study the temperature induced coalescence of two tubes. 
The presence of defects in the tubes may act as a driving force for 
coalescence.
It has been shown that vacancies are the leading defects promoting
coalescence\cite{terr00}.
Vacancies can be produced by chemical treatment\cite{nik97}, electron 
irradiation\cite{terr00}, etc.
Certainly, the thermal treatment of the tubes at relatively high 
temperatures (above 1600$^{\rm o}$C) will produce the required vacancies
in the tubes.
Due to the high temperatures, the vacancies have a high mobility and 
diffuse throughout the tube until they get locked in the intertube region 
by saturation of the associated dangling bonds with the dangling bonds 
left free by a nearby vacancy in the adjacent tube. 
Saturation of dangling bonds between neighboring tubes yields to 
intertube polymerization which is the initial stage for the coalescence
of tubes.
At the experimental temperatures, the creation of vacancies in the tubes
requires heat treatments of about minutes. 
This time scale is clearly not tractable in conventional MD simulations.
However, once a sufficient number of intertube links are formed, coalescence
proceeds within a time scale of a few hundred picoseconds.
It is this second part of the process which is suitable for being simulated
in the computer and therefore we concentrate our study in this part.

\begin{figure}[t]
   \begin{center}
   \begin{tabular}{c}
   \includegraphics[height=12cm,angle=270]{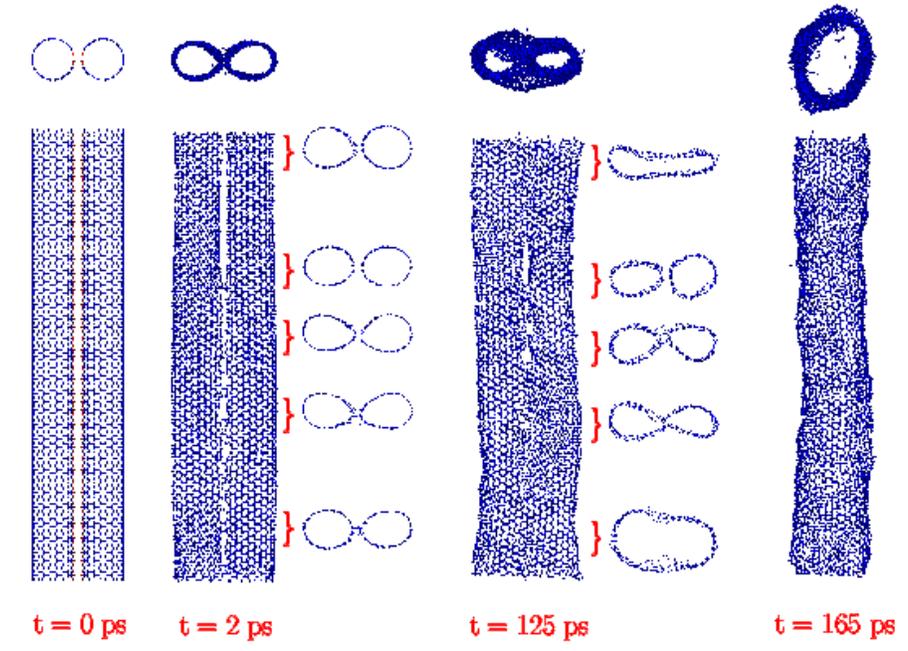}
   \end{tabular}
   \end{center}
   \caption
   { \label{figSWMW1}
   Snapshots (side and top views) showing the sequence of the coalescence
   of two (10,10) tubes. Also shown top views of some tube portions
   embraced by brackets.}
   \end{figure}

We have simulated the coalescence of two (10,10) tubes under thermal 
treatment (see Fig. \ref{figSWMW1}).
The initial intertube distance between the tubes is of 17 {\rm \AA}.
The first step consists in the generation of vacancies by random
removal of a number (2\%) of atoms.
Since no significant diffusion of vacancies is expected within the
time scale of the simulations, only those vacancies in the vicinity
of the partner tube play an active role in the coalescence process.
Therefore we restrict the removal of atoms to the intertube region.
Moreover, long simulation cells, containing 60 unit cells in the axial 
direction of the tubes, are considered to avoid unphysical uniform 
distributions of vacancies arising from the use of periodic boundary 
conditions in the axial direction of the tubes.
Polymerization of the two tubes occurs very rapidly (in 2 ps) after
the generation of vacancies.
The formation of intertube links is driven by the saturation of the dangling
bonds left free by the vacancies which results in a non uniform 
distribution of the links along the tubes length (see Fig. \ref{figSWMW1},
$t$=2 ps).
By progressively heating up the tubes, the initial intertube links 
develop into enveloping surfaces, or ``patches", common to the two 
tubes.
The ``tearing" apart of the intratube bonds in the intertube region in
some of the patches gives rise to partial coalescence of the tubes
whereas other portions of the tubes remain polymerized or even 
completely unlinked (see Fig. \ref{figSWMW1}, $t$=125 ps).
As the time evolves, new intertube links and patches develop,
the existing patches and the coalesced parts of the tubes grow in 
the axial direction and merge one to another, till, eventually,
all the remaining intratube bonds in the intertube region
``tear" apart producing the coalesce of the two tubes in their whole length
(see Fig. \ref{figSWMW1}, $t$=165 ps).
The patching--and--tearing mechanism emerging from our simulations,
explains the coalescence of two tubes in great detail. 
This mechanism gives support to the early intuitive ``zipping"
description of coalescence\cite{nik97}.
Moreover, our results on the coalescence of two (10,10) tubes, are in fair 
agreement with the tight-binding simulations of Terrones 
{\em et al.} \cite{terr00}.

   \begin{figure}[t]
   \begin{center}
   \begin{tabular}{c}
   \includegraphics[height=10cm]{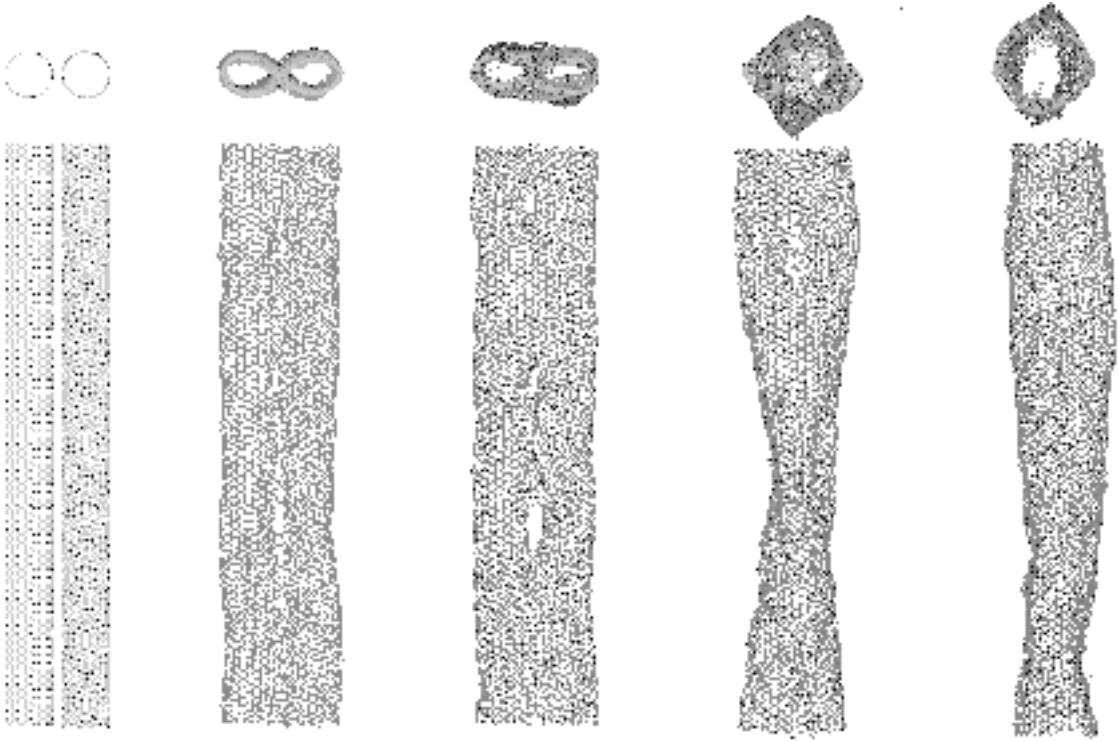}
   \end{tabular}
   \end{center}
   \caption
   { \label{figSWMW2}
   Snapshots (side and top views) showing the sequence of the coalescence
   of a (10,10) and a (12,8) tubes.}
   \end{figure}
We have also found, in contrast to the earlier statement\cite{terr00} that
this process was very unlikely, that coalescence proceeds between tubes 
of different chiralities without any restriction.
Figure \ref{figSWMW2} shows the coalescence of a (10,10) and a (12,8) tubes.
The two tubes have similar radii and a long simulation cell containing
approximately (due to the incommensurability of the two tubes)
68 and 9 unit cells, respectively, has been used in the simulations.
We find coalescence of the tubes even in the most stringent case 
of two non chiral 
tubes one of the armchair type, the (10,10) tube, and the other one of 
the zigzag type, the (17,0) tube.
For both pairs of tubes, large tube reconstructions are required for 
knitting the two corresponding tubes into a larger diameter SWCNT, 
due to the mismatch of their respective honeycomb structures.
The extent of reconstruction required can only be accomplished by using long
simulation cells as the ones used in the two examples presented in
this paper.
In summary, we have demonstrated that no restrictions apply for the 
coalescence of tubes of different chiralities, provided that the simulation 
cell in the axial direction of the tubes is long enough.
Coalescence proceeds following the same patching--and--tearing mechanism
introduced to explain coalescence of non chiral tubes.
However, it is fair to recognize that the resulting tubes have
higher concentrations of defects than in the case of coalescence of non 
chiral tubes of the same type.
Little or no annealing of those defects is observed within the limited
time scale of our simulations.

As we have pointed out above, large simulation cells are used to avoid
unphysical uniform distributions of vacancies arising from the repetition
of the cell within the periodic boundary conditions scheme and to accommodate 
the tube reconstruction involved in the coalescence of tubes with different 
chiralities.
However, the use of smaller cells would be preferred from the computational
(time saving) view point.
We have studied the effect of reducing  the size of the simulation cell
down to 20 unit cells in the axial direction of the tubes, in the case
of coalescence of two non-chiral (10,10) tubes.
The use of the smaller cell does not change the  outcome of the
simulated coalescence and the patching--and--tearing mechanism 
describing the process remains valid.
Therefore in the next section, where we simulate bundles of (10,10)
tubes, we will use the smaller simulation cell without any loss of
generality or physical insight.

\subsection{Transformation of SWCNT's ropes into MWCNT's}
\label{subsect:SWtoMW}

\begin{figure}[h]
   \begin{center}
   \begin{tabular}{c}
   \includegraphics[height=10cm]{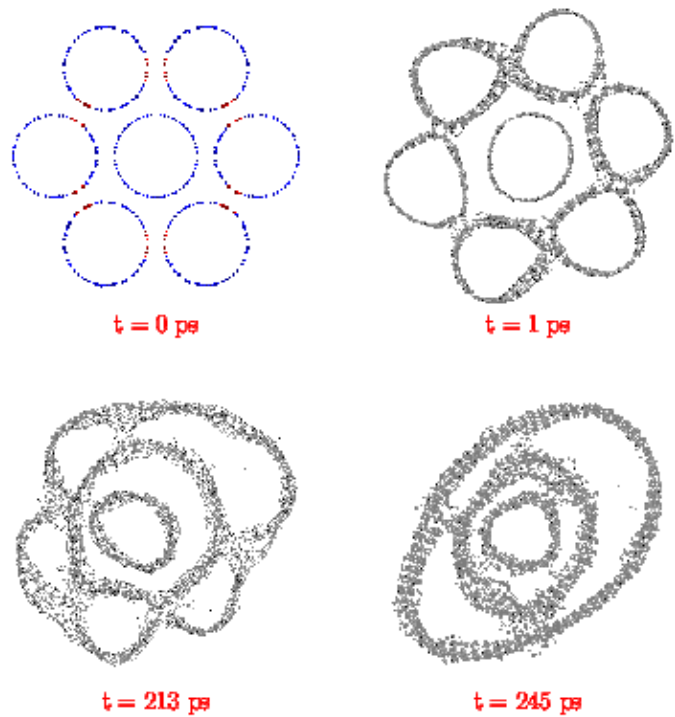}
   \end{tabular}
   \end{center}
   \caption
   { \label{figSWMW4}
   Snapshots (top views) taken from the simulations showing the sequence
   of the SW-to-MW transformation of a bundle of seven (10,10) tubes.}
   \end{figure}

Let us first consider a bundle of seven (10,10) tubes arranged in a 
triangular lattice (with a lattice parameter of 17 {\rm \AA})
with a central tube and six surrounding tubes
(see Fig. \ref{figSWMW4}). 
For investigating the temperature induced SW-to-MW transformation
we assume that, similarly to the case of coalescence of two tubes, 
the vacancies play the leading role in promoting the structural
reorganization of the bundles.
This assumption will be lately confirmed by the simulations (see below).
The thermal treatment of the bundle at elevated temperatures will produce 
a number of vacancies in the tubes.
Therefore we begin the simulation by generating a number (4\%) of vacancies 
in the intertube regions defined by the six outer tubes
(no defects are created in  the central tube, see below for a justification).
Upon the generation of vacancies, polymerization between neighboring tubes
takes place in less than 1 ps (see Fig. \ref{figSWMW4}, $t$=1 ps).
The bundle is then progressively heated up.
Coalescence between neighboring tubes begins to develop following the 
patching-and-tearing mechanism described above (see Fig. \ref{figSWMW4}, 
$t$=213 ps).
The concerted coalescence of the tubes around the bundle 
gives rise to the transformation of the bundle of SWCNTs into a MWCNT 
consisting of three nested tubes (see Fig. \ref{figSWMW4}, $t$=245 ps).
The central tube may also have a number of defects. Then, depending on the
links established between the central tube and its surroundings, three (as
the one shown here) or two nested structures may occur.

\begin{figure}[t]
   \begin{center}
   \begin{tabular}{c}
   \includegraphics[height=9cm]{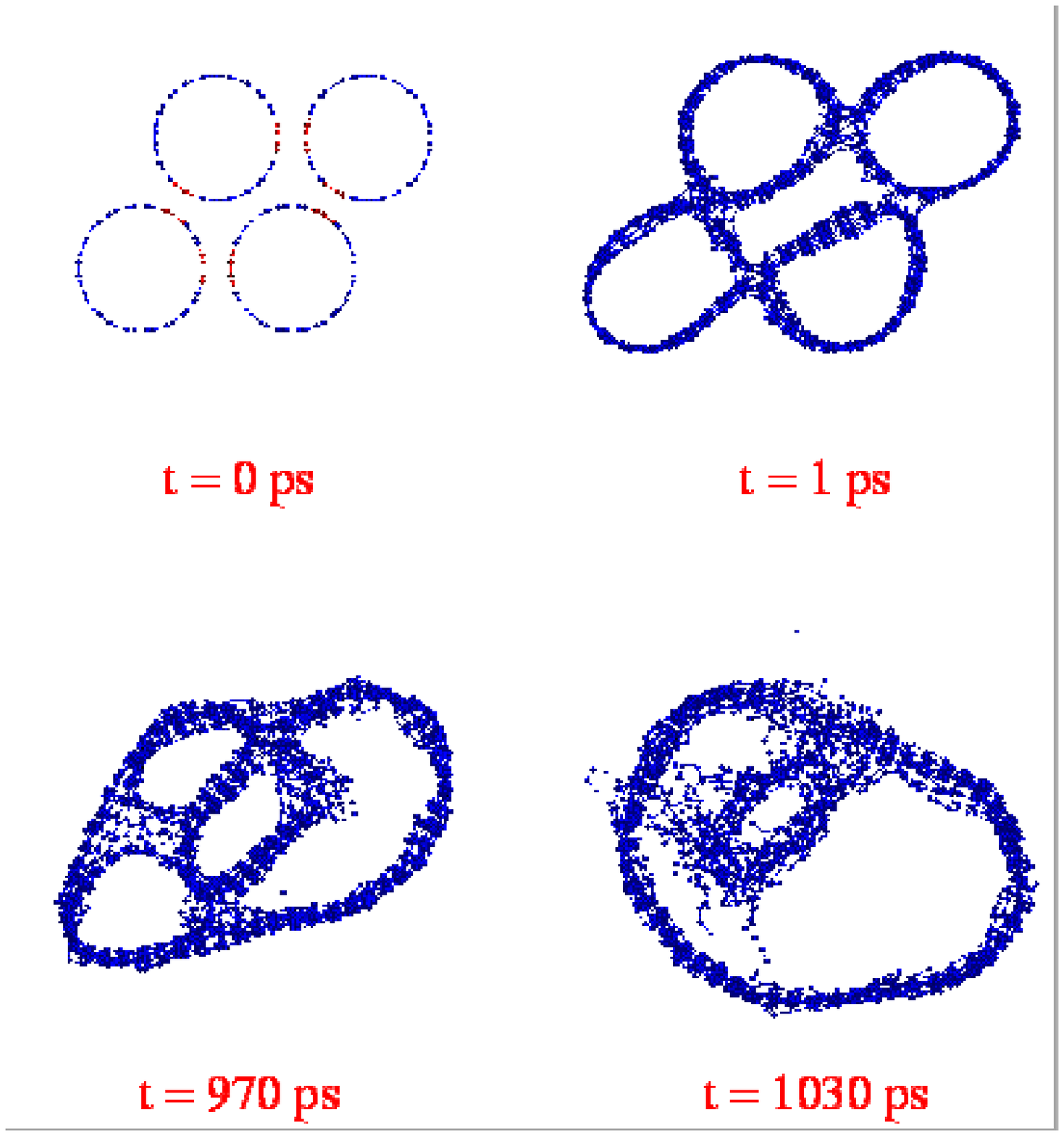}
   \end{tabular}
   \end{center}
   \caption
   { \label{figSWMW5}
   Snapshots (top views) taken from the simulations showing the 
   transformation of a bundle of four (10,10) tubes into a MWCNT
   consisting on two nested tubes.}
   \end{figure}

Bundles containing four and five tubes transform into MWCNTs
consisting of two nested tubes (see Fig. \ref{figSWMW5})
following a similar patching-and-tearing mechanism operating
in a concerted way around the bundle.

   \begin{figure}[h]
   \begin{center}
   \begin{tabular}{c}
   \includegraphics[height=8cm]{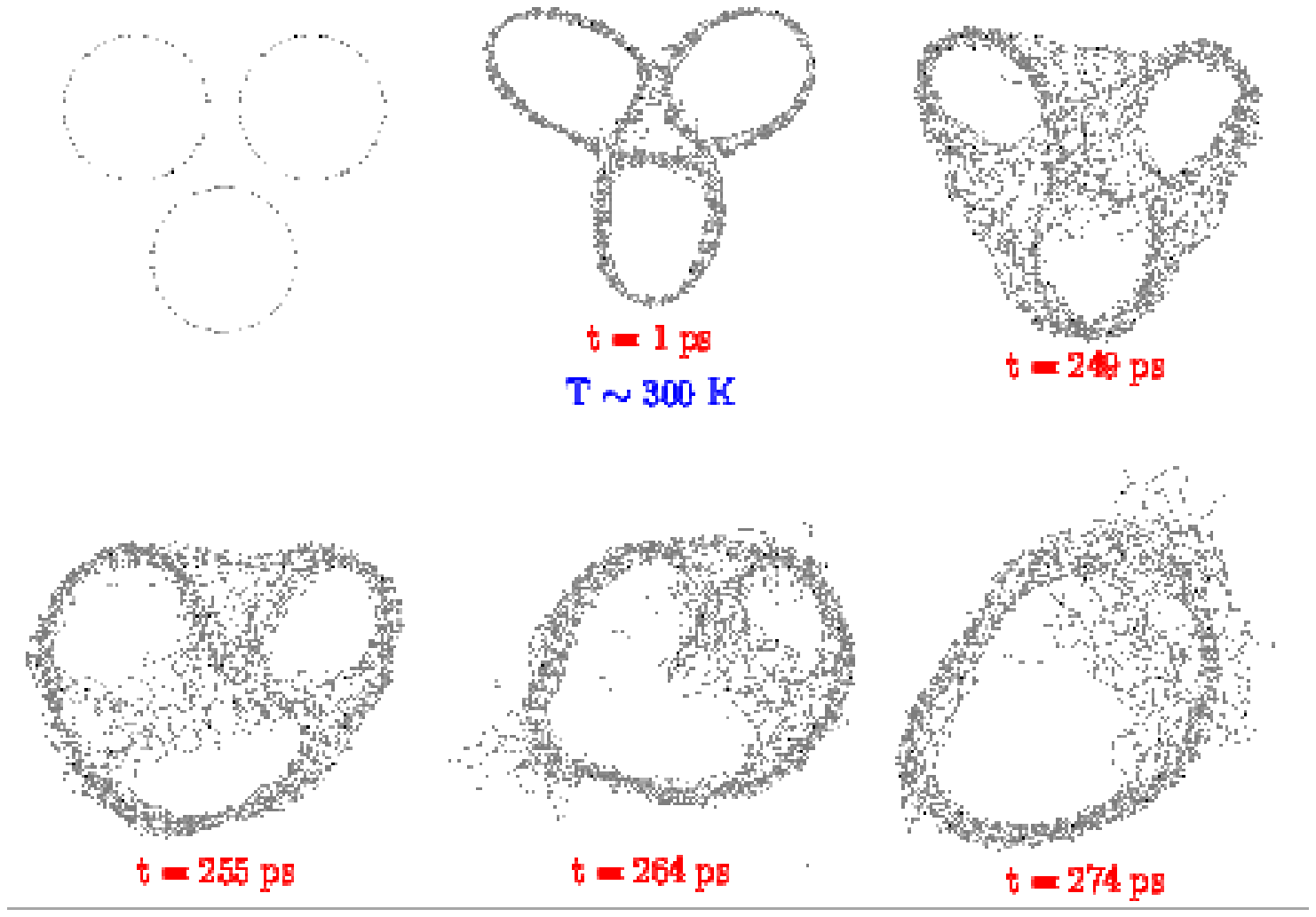}
   \end{tabular}
   \end{center}
   \caption
   { \label{figSWMW6}
   Snapshots (top views) taken from the simulations showing the 
   coalescence of a bundle of three (10,10) tubes into a larger
   diameter SWCNT.}
   \end{figure}
We find that four tubes in the bundle is the lower limit to observe
the SW-to-MW transformation.
Thus, the concerted coalescence of three (10,10) SWCNTs produces 
a SWCNT of larger diameter (see Fig. \ref{figSWMW6}).
One could argue about the small amount of C atoms available to 
reconstruct an internal tube.
However, we find that a bundle of three (20,20) tubes (where the supply 
of C atoms would be sufficient to reconstruct an internal (9,9) tube) also
coalesces in a SWCNT of larger diameter.

The excellent agreement between the simulations and the experiments
gives support for the assumptions made about the role played by the
vacancies in the SW-to-MW transformation.
We want to stress also that the patching-and-tearing mechanism 
presented here is of general applicability to describe both
the coalescence of SWCNTs into a larger diameter tube and 
the newest SW-to-MW transformation based on the concerted coalescence 
of the tubes in a bundle.

\section{Summary}
\label{sect:summary}

Novel crystalline ropes of polygonized SWCNTs (tube diameter $D$ about 
17 {\rm \AA}) produced by CO$_2$ laser ablation exhibit rounded-hexagonal 
cross sections.
Simulations of the lattice show several metastable structures 
characterized by different tube cross sections, 
hexagonal, rounded-hexagonal and circular, and increasing cell volume. 
The lowest energy configuration of the lattice progresses from
circular to hexagonal with increasing tube diameter.
The driving force for the departure of the tubes from the ideal circular
shape is the attractive (van der Waals-type) intertube interaction.
In contrast to the experimental ropes, the lowest energy configuration
of the lattice of tubes with $D=17$ {\rm \AA} corresponds to nearly
circular tubes. 
However, the lattice of rounded-hexagonal tubes is very close in 
energy, what would explain the experimental observation.
It would be of great interest to understand and control the production
of the different metastable structures of the lattice depending
on the growth conditions.

The experimental observation of the transformation of SWCNT bundles
into MWCNTs under high temperature treatments motivated us
to simulate the thermal stability of the ropes.
Extensive Molecular Dynamics simulations combined with the experimental
results unveil the physical mechanism of the SW-to-MW transformation
based on the patching of SWCNTs and their subsequent tearing apart
giving rise to a MWCNT.
The driving force for this transformation are the vacancies created
by the high temperature treatment.
Further studies are required to improve our understanding of the 
structural stability of the ropes and their possible morphological
changes.
Possible implications on the different growth regimes of SWCNT 
bundles are still obscure.

\acknowledgments 
The work has been done in collaboration with the groups of
S. Iijima (Japon) and Sylvie Bonamy (Orelans, France).
Work supported by DGI of Spain (Grant MAT2002-04499-C02),
European Community (RTN-COMELCAN), and
Junta de Castilla y Le\'on (Grant CO01/102).
M.J.L. acknowledges support from the Spanish MCYT under the
``Ram\'on y Cajal" Program.




\end{document}